\documentstyle[prd,aps,floats,epsfig]{revtex}
\tighten
\begin{document}
\draft
\twocolumn[\hsize\textwidth\columnwidth\hsize\csname 
@twocolumnfalse\endcsname
\title{GLOBAL VS LOCAL COSMIC STRINGS FROM PSEUDO-ANOMALOUS U(1).}
\author{P.~Bin\'etruy$^a$, C.~Deffayet$^a$, P.~Peter$^b$}
\address{$^a$LPTHE, Universit\'e Paris-XI, B\^atiment 211, F-91405
Orsay Cedex, France\\ $^b$DARC, Observatoire de Paris-Meudon, UPR 176,
CNRS, F-92195 Meudon, France}
\date{19 June 1998}
\preprint{LPTHE-ORSAY/9840}
\maketitle
\begin{abstract}
We study the structure of cosmic strings produced at the breaking of
an anomalous U(1) gauge symmetry present in many superstring
compactification models. We show that their coupling with the axion
necessary in order to cancel the anomalies does not prevent them from
being local, even though their energy per unit length is found to
diverge logarithmically. We discuss the formation of such strings and
the phenomenological constraints that apply to their parameters.
\end{abstract}
\vskip2pc]

\section*{Introduction}

Most classes of superstring compactification lead to a spontaneous
breaking of a pseudo-anomalous U(1) gauge symmetry~\cite{U1} whose
possible cosmological implications in terms of inflation scenarios
were investigated~\cite{casas}-\cite{ERR}. Anomalies are cancelled
through a mechanism which is a four-dimensional version of the famous
anomaly cancellation mechanism of Green and Schwarz \cite{GS} in the
10-dimensional underlying theory; the coupling of the dilaton-axion
field to the gauge fields plays a central role.  Cosmic
strings~\cite{kibble} might also be formed in this framework; because
of their being coupled to the axion field, such strings were thought
to be of the global kind~\cite{casas,harvey,jeannerot,LR}. It is our
purpose here to show that there exists a possibility that (at least
some of) the strings formed at the breaking of this anomalous U(1) be
local, in the sense that their energy per unit length can be localized
in a finite region surrounding the string's core, even though this
energy is formally logarithmically infinite. It will be shown indeed
that the axion field configuration can be made to wind around the
strings so that any divergence must come from the region near the core
instead of asymptotically. The cutoff scale that then needs be
introduced is thus a purely local quantity, definable in terms of the
microscopic underlying fields and parameters. It is even arguable that
such a cutoff should be interpreted as the scale at which the
effective model used throughout ceases to be valid.

\section{The model}

The antisymmetric tensor field $B_{\mu\nu}$ which appears among the
massless modes of the closed string plays a fundamental role in the
cancellation of the anomalies. Indeed Green and Schwarz~\cite{GS} had
to introduce a counterterm of the form $B \wedge F \wedge F \wedge F
\wedge F$ in order to cancel the gauge anomalies of the 10-dimensional
theory, $F$ being the gauge field strength. Compactifying six
dimensions to get to four spacetime dimensions and replacing three of
the field strength with compact indices by their background values
yields a term proportional to $B \wedge F$.  It is well-known that in
four spacetime dimensions an antisymmetric tensor is dual to a
pseudoscalar $a$: $\epsilon_{\mu\nu\lambda\tau}
\partial^{\nu} B^{\lambda \tau} \sim \partial_\mu a$. The coupling is
then simply $A^\mu \partial_\mu a$.

In this formulation, the pseudoscalar belongs to the same (chiral)
supermultiplet as the dilaton field $s$ and they form a complex scalar
field $S=s+ia$.  This superfield couples in a model independent way to
the gauge fields present in the theory; one has in particular:
\begin{equation}
{\cal L} = -{s \over 4 M_{_P}} \sum_a F^a_{\mu\nu} F^a_{\mu\nu}
           +{a \over 4 M_{_P}} \sum_a F^a_{\mu\nu} \widetilde
F^a_{\mu\nu},
\end{equation}
where $M_{_P}$ is the reduced Planck scale, the index $a$ runs over
all gauge groups and
\begin{equation} \widetilde F^{a\mu\nu}\equiv {1\over 2}
\varepsilon^{\mu\nu\rho\sigma}
F^a_{\rho\sigma}.\label{deftilde}\end{equation} Thus $a$ has
axion-like couplings and is indeed called the string axion.  And the
vacuum expectation value of the dilaton $\langle s\rangle $ yields the
gauge coupling $1/g^2$.

An abelian symmetry with gauge field $A_\mu$ may seem to have (mixed)
anomalies: under $A_\mu \rightarrow A_\mu + \partial_\mu \alpha$ $$
\delta {\cal L} = -{1 \over 2} \delta_{_{GS}} \alpha \sum_a
F^a_{\mu\nu}
\widetilde F^a_{\mu\nu}.$$
But this can be cancelled by an apropriate shift of the string axion
$a$.  Since there is a single model-independent axion, only one
abelian symmetry, henceforth referred to as $U(1)_X$, may be
pseudo-anomalous.

The kinetic term for the dilaton-axion fields is described at the
supersymmetric level by the D-term of the K\"ahler function $K= -
\ln (S + \bar S)$. This may be modified to include the Green-Schwarz 
term $A^\mu \partial_\mu a$~\cite{U1}: $K= -\ln (S + \bar S -
4\delta_{_{GS}} V)$ where $V$ is a vector superfield describing the
anomalous $U(1)_X$ vector supermultiplet. This D-term includes other
terms, such as a mass term for the $A_\mu$ gauge field. The constant
$\delta_{_{GS}}$ may be computed in the framework of the weakly
coupled string and is found to be~\cite{U1}:
\begin{equation}
\delta_{_{GS}} = {1 \over 192 \pi^2} \sum_i X_i,
\end{equation}
where $X_i$ are the charges of the different fields under $U(1)_X$.

We are therefore led to consider the general class of models described
by the Lagrangian (restricted here to the bosonic fields)
\begin{eqnarray} {\cal L} &=& 
-(D_{\mu}\Phi_i )^\dagger (D^{\mu}\Phi_i)\nonumber\\
& & -{1\over 4g^2} \left( F_{\mu\nu} F^{\mu\nu} -{a\over
M_{_P}}F^{\mu\nu}
\widetilde F_{\mu\nu}\right)\nonumber \\
& & -\delta_{_{GS}}^2  M_{_P}^2 A_\mu A^\mu +
\delta_{_{GS}}  M_{_P} A^\mu \partial _\mu a \nonumber\\
& & -{1\over 4} \partial _\mu a \partial^\mu a -{1\over 4} m^2
a^2-V(\Phi_i) ,\label{lag}\end{eqnarray} where we have set the dilaton
field to its vacuum expectation value $\langle s\rangle = 1/g^2$, we
have included a mass term for the axion, without specifying its
origin, and we have introduced scalar fields $\Phi_i$ carrying the
integer charge $X_i$ under the $U(1)_X$ symmetry; the covariant
derivative is defined by
\begin{equation}
D^{\mu}\Phi_i \equiv (\partial^\mu - iX_i A^{\mu})\Phi_i
,\label{defderiv}\end{equation}
and the potential $V(\Phi_i)$ by
\begin{equation}
V(\Phi_i) \equiv {g^2\over 2} (\Phi_i^\dagger X_i \Phi_i +\delta_{_{GS}}
M_{_P}^2)^2 ,\label{pot}
\end{equation}
The Green-Schwarz coefficient $\delta_{_{GS}}$ and the axion field $a$
have been rescaled by a factor $g^2$.

The lagrangian ~(\ref{lag}) is invariant under the following local
gauge transformation with gauge parameter $\alpha(x^\mu)$
\begin {eqnarray} \Phi_i &\rightarrow& \Phi_i e^{iX_i\alpha} \nonumber
\\
A_{\mu} &\rightarrow& A_{\mu} + \partial_{\mu}\alpha \label{gauge}\\
a&\rightarrow&a+2M_{_P}\delta_{_{GS}}\alpha,\nonumber \end {eqnarray}
the transformation of the term $(a/ 4g^2 M_{_P})F^{\mu\nu}
\widetilde F_{\mu\nu}$ cancels the variation of the effective 
lagrangian due to the anomaly, namely $\delta {\cal L} = -(1/2g^2
)\delta_{_{GS}}\alpha F^{\mu\nu} \widetilde F_{\mu\nu}$ (assuming we
are also transforming the fermions of the theory not written
explicitely in ~(\ref{lag})).  Making a rigid gauge transformation
with parameter $\alpha = 2\pi$ without changing $a$ as a first step
but transforming the other fields (including the fermions), leads us
to interpret $a$ as a periodic field of period $4\pi
\delta_{_{GS}}M_{_P}$ through the redefinition $a
\rightarrow a-4 \pi \delta_{_{GS}}M_{_P}$ which leaves
the lagrangian invariant. It is also manifest in the following
rewriting of the kinetic term and of the axionic $\theta$-term in
${\cal L}$ where it is clear that $a$ behaves like a phase:

\begin {eqnarray}
{\cal L}_{kin,\theta}&=&-\frac{1}{4g^2}\left(F^{\mu\nu}F_{\mu\nu} -
\frac{a}{M_{_P}}F^{\mu\nu}\widetilde F_{\mu\nu}\right) \nonumber\\
& &  -\partial^{\mu}\phi_i\partial_{\mu}\phi_i -
\phi_i^2X_i^2\left(\frac{\partial^{\mu}\eta_i}{X_i}-A^{\mu}\right)^2\nonumber\\
& & - {M_{_P}}^2\delta_{_{GS}}^2\left(\frac{\partial^{\mu}a}
{2M_{_P}\delta_{_{GS}}}-A^{\mu}\right)^2
\end {eqnarray}
where we have set $\Phi_i\equiv\phi_{i}e^{i\eta_i}$ ($\phi_i$ being
the modulus of $\Phi_i$).\\ Let us now work out the Higgs mechanism in
this context. We consider for the sake of simplicity a single scalar
field $\Phi$ of negative charge $X$ and we drop consequently the $i$
indices.  ${\cal L}_{kin,\theta}$ can be rewritten

\begin{eqnarray}
\lefteqn{{\cal L}_{kin,\theta}=-\left[ M_{_P}^2\delta^2_{_{GS}} 
+ \phi^{2}X^{2}\right]\times}\nonumber
\\
& &\left[A^\mu-\frac
 {\frac{1}{2}M_{_P}\delta_{_{GS}}\partial^\mu a+\phi^2X\partial^\mu
\eta}{M_{_P}^2\delta_{_{GS}}^2+\phi^2X^2}\right]^2\nonumber\\
&& -\frac{\phi^2M_{_P}^2\delta^2_{_{GS}}X^2}{
M_{_P}^2\delta_{_{GS}}^2 + \phi^2X^2}\left[\frac
{\partial^{\mu}a}{2M_{_P}\delta_{_{GS}}}-\frac
{\partial^\mu\eta}{X}\right]^2 \nonumber\\
&& +\frac{\delta_{_{GS}}}{2g^2}\left(\frac{\phi^2X^2}{M_{_P}^2
\delta_{_{GS}}^2+\phi^2X^2}\left[\frac{a}{2M_{_P}\delta_{_{GS}}}
-\frac{\eta}{X}\right]\right.
\nonumber\\
&&\left.+ \frac {\frac{1}{2}M_{_P}\delta_{_{GS}}a+\phi^2X \eta}
{M_{_P}^2\delta_{_{GS}}^2+\phi^2X^2}\right) F_{\mu\nu}\widetilde
F^{\mu\nu}
-\frac{1}{4g^2}F_{\mu\nu}F^{\mu\nu} .
\end{eqnarray}

The linear combination appearing in this last equation
\begin{equation}
\frac{a}{2M_{_P}\delta_{_{GS}}}-\frac{\eta}{X}
\end{equation}
is the only gauge invariant linear combination of $\eta$ and $a$ (up
to a constant).  The other one
\begin{equation}
\ell \equiv \frac {\frac{1}{2}M_{_P}\delta_{_{GS}}a+\phi^2X \eta}
{M_{_P}^2\delta_{_{GS}}^2+\phi^2X^2}
\end{equation}
has the property of being linearly independent of the previous one and
of transforming under a gauge transformation ~(\ref{gauge}) as $\ell
\rightarrow \ell + \alpha$.  We now assume explicitely that $\Phi$
takes its vacuum expectation value $\langle \Phi^\dagger \Phi
\rangle \equiv \rho^2$ in order to minimize the potential~(\ref{pot}):
\begin{equation}
\rho^2=-\delta_{_{GS}}M_{_P}^2/X.\label{12}
\end{equation}
We are left, among other fields, with a massive scalar Higgs field
corresponding to the modulus of $\Phi$ of mass $m_{_X}$ given by
\begin{equation}
m_{_X}^2 = 2g^2\rho^2X^2=-2\delta_{_{GS}}Xg^2M_{_P}^2
\end{equation}
and we define
\begin{equation}
\hat a \equiv \left[\frac{a}{2M_{_P}\delta_{_{GS}}}-
\frac{\eta}{X}\right]\frac{\sqrt{2}\rho M_{_P}\delta_{_{GS}}X}
{\sqrt{M_{_P}^2\delta_{_{GS}}^2+\rho^2 X^2}}\label{hata}
\end{equation}
and 
\begin{eqnarray}
F_{a}^2 &= &\frac{1}{128\pi^4}\frac{M_{_P}^2 g^{4}}{\rho^2 X^2}
\left( M_{_P}^2 \delta^2_{_{GS}}+\rho^2X^2\right) \nonumber\\
 &=&\frac{1}{128\pi^4}M_{_P}^2 g^4
\left(1+\left(\frac{m_{_X}}{M_{_P}}\right)^2\frac{1}{2g^2X^2}\right)
\end{eqnarray}
so that with $\rho$ being set:
\begin{eqnarray}
\lefteqn{{\cal L}_{kin,\theta} = \nonumber} \\
&& -\left[ M_{_P}^2\delta_{_{GS}}^2
+\rho^2X^2\right]\left[A^{\mu} -\partial ^\mu \ell\right]^2
-\frac{1}{2}\partial^{\mu}\hat a \partial_{\mu} \hat a \nonumber \\
&&+\left[\frac{\hat
a}{32\pi^2F_{a}}+\frac{\delta_{_{GS}}}{2g^2}\ell\right]F_{\mu\nu}
\widetilde F^{\mu\nu}-\frac{1}{4g^2}F_{\mu\nu}F^{\mu\nu}
\end{eqnarray}
we can now make a gauge transformation to cancel $\partial ^\mu \ell $
by setting $\alpha = -\ell + \beta$ where $\beta$ is a constant
parameter.  This leaves us with
\begin{eqnarray}
\lefteqn{{\cal L}_{kin,\theta}= -\frac{m^2_{_A}}{2g^2}A^{\mu}
A_{\mu}-\frac{1}{2}\partial^{\mu}\hat a \partial_{\mu} \hat a
}\nonumber\\ &&+\frac{\hat a}{32\pi^2F_{a}}F_{\mu\nu}
\widetilde F^{\mu\nu}-\frac{1}{4g^2}F_{\mu\nu}F^{\mu\nu} ,
\end{eqnarray} 
where $m_{_A}$ given by
\begin{eqnarray}
m_{_A}^2 &=& 2g^2\left[\rho^2X^2+M^2_{_P}\delta_{_{GS}}^2\right]\nonumber \\
&= &m_{_X}^2 \left[
1+\left(\frac{m_{_X}}{M_{_P}}\right)^2\frac{1}{2g^2X^2}
\right] \label{mA}
\end{eqnarray}
is the mass of the gauge field after the symmetry breaking.
The remaining symmetry
\begin{equation}
\hat a \rightarrow 
\hat a  + \frac {32\pi^2F_a}{2g^2}\delta_{_{GS}}\beta
\end{equation}
is the rigid Peccei-Quinn symmetry which compensates for the anomalous
term arising from a rigid phase transformation of parameter $\beta$.

To summarize we have seen that in the presence of the axion the gauge
boson of the pseudo-anomalous symmetry absorbs a linear combination
$\ell$ of the axion and of the phase of the Higgs field. We are left
with a rigid Peccei-Quinn symmetry, the remnant axion being the other
linear combination $\hat a$ of the original string axion and of the
phase of the Higgs field.

\section{Pseudo-anomalous U(1) strings}

Cosmic strings can be found as solutions of the field equations
derivable from Eq.~(\ref{lag}) provided the underlying U(1) symmetry
is indeed broken, which implies that at least one of the eigenvalues
$X_i$ is negative. This is the first case we shall consider here, so
we shall in this section assume again only one field $\Phi$ with
charge $X$, with $X<0$. Assuming a Nielsen-Olesen-like solution along
the $z-$axis~\cite{NO}, we set, in cylindrical coordinates,
\begin{equation} \Phi = \phi (r) \hbox{e}^{i\eta}, \ \ \ \ \eta =
n\theta, \end{equation} for a string with winding number $n$. This
yields the following Euler-Lagrange equations
\begin{equation} \Box a = 2\delta_{_{GS}} M_{_P} \partial_\mu
A^\mu -{1\over 2g^2 M_{_P}}F_{\mu\nu}\widetilde F^{\mu\nu} + m^2 a,
\label{Boxa}\end{equation}
\begin{equation} \Box \phi = \phi (\partial _\mu\eta-XA_\mu)^2 
+g^2X\phi (X\phi^2 +\delta_{_{GS}} 
M_{_P}^2),\label{Boxphi}\end{equation}
\begin{equation} \partial_\mu [\phi^2 (\partial^\mu \eta - X A^\mu)]
= 0,\label{Boxeta}\end{equation}
\begin{eqnarray}{1\over g^2} \partial_\mu ({a\over M_{_P}}\widetilde 
F^{\mu\nu} - F^{\mu\nu})&=&\delta_{_{GS}} M_{_P} \partial ^\nu a
-2\delta_{_{GS}}^2 M_{_P}^2 A^\nu \nonumber\\ & & +2X\phi^2
(\partial^\nu\eta-XA^\nu),\label{BoxA}\end{eqnarray} from which the
string properties can be derived.

Eq.~(\ref{BoxA}) can be greatly simplified: first we make use of
Eq.~(\ref{deftilde}), which implies $\partial_\mu \widetilde
F^{\mu\nu} =0$, and then we derive Eq.~(\ref{BoxA}) with respect to
$x^\nu$. This gives, upon using Eqs.~(\ref{Boxa}) and (\ref{Boxeta}),
\begin{equation} F_{\mu\nu}\widetilde F^{\mu\nu} = 2m^2 M_{_P} g^2
a,\label{FFa}\end{equation}
and, with the help of Eq.~(\ref{BoxA}),
\begin{equation} {1\over g^2}\partial_\mu F^{\mu\nu} = {1\over M_{_P}}
\widetilde F^{\mu\nu}\partial_\mu a + {\cal J}^\nu +
J^\nu,\label{BoxA2} \end{equation}
where the currents are defined as
\begin{equation} J^\mu = -2 X\phi^2 (\partial^\mu\eta -
XA^\mu),\label{J}\end{equation}
and
\begin{equation}{\cal J}^\mu = -\delta_{_{GS}}  M_{_P}
(\partial ^\mu a-2\delta_{_{GS}} M_{_P}A^\mu).
\label{calJ}\end{equation}
Eqs.~(\ref{Boxa}) and (\ref{Boxeta}) then simply express those two
currents conservation $\partial \cdot J = \partial \cdot {\cal J} =0$,
when account is taken of Eq.~(\ref{FFa}).

The standard paradigm concerning the strings obtained in this simple
model states that the presence of the axion makes the string global in
the following sense: even for a vanishing $a$, $A_\mu$ behaves
asymptotically in such a way as to compensate for the Higgs field
energy density (i.e., $A_\mu \to -\partial_\mu\eta/X$) and therefore
yields an energy per unit length which diverges asymptotically. It
should be clear however that the behaviour of $a$ could be different;
indeed, it could as well compensate for this divergence as we will now
show. In this case then, a divergence is still to be found, but this
time at a small distance near the string core, so that the total
energy is localised in a finite region of space.  This is in striking
contrast with the case of a global string where the divergent behavior
arises because the energy is not localized and a large distance
cut-off must be introduced.

In order to examine the behaviour of the fields and the required
asymptotics, we need the stress energy tensor
\begin{equation} T^\mu_\nu = -2g^{\mu\gamma} {\delta {\cal L}\over
\delta g^{\gamma\nu}}+\delta^\mu_\nu {\cal L},\end{equation}
which reads explicitely
\begin{eqnarray} T^{\mu\nu}&=&2[\partial^\mu\phi\partial^\nu\phi
-{1\over 2}g^{\mu\nu}(\partial \phi)^2]\nonumber \\ & & +{1\over g^2}
(F^{\rho\mu} F_\rho ^{\ \nu}-{1\over 4}g^{\mu\nu}F\cdot F)
\\
& & -{1\over 2} g^2 g^{\mu\nu} (X\phi^2+
\delta_{_{GS}}M_{_P}^2 )^2\nonumber\\
& & +{1\over 2\delta_{_{GS}}^2 M_{_P}^2 } [{\cal J}^\mu {\cal
J}^\nu -{1\over 2} g^{\mu\nu} {\cal J}^2]\nonumber\\ & & +{1\over 2
X^2
\phi^2} [J^\mu J^\nu -{1\over 2} g^{\mu\nu} J^2]-{1\over 4} m^2 a^2
g^{\mu\nu}
\nonumber\\
\label{Tmunu}\end{eqnarray}
where account has been taken of the field equations. The energy per
unit length $U$ and tension $T$ will then be defined respectively as
\begin{equation} 
U=\int d\theta\, rdr T^{tt} \ \ \hbox{ and }\ \ 
T=-\int d\theta\,rdr T^{zz},\label{UT}
\end{equation}
The question as to whether the corresponding string solution is local
or global is then equivalent to asking whether these quantities are
asymptotically convergent ({\em i.e.} at large distances).

It can be seen on Eq.~(\ref{Tmunu}) that only the last two terms can
be a potential source of divergences. The Nielsen-Olesen~\cite{NO}
solution for the very last term consists in saying that $A_\mu$ is
pure gauge, namely $\lim_{r\to\infty} D_\mu \Phi = 0$, so that, as
already argued, $\lim_{r\to\infty} A_\mu = -\partial_\mu\eta/X$. With
this solution, setting $a=0$ implies that the second to last term in
Eq.~(\ref{Tmunu}) should diverge logarithmically for
$r\to\infty$. However, at this point, it should be remembered that $a$
can be interpreted as a periodic field of period $4\pi
\delta_{_{GS}}M_{_P}$ (as long as a cosine-like mass term is not
included as is usually the case at very low temperatures if this axion
is to solve the strong CP problem of QCD) and therefore can be
assigned a variation along $\eta$. In fact, setting
\begin{equation} a = {2 \delta_{_{GS}}M_{_P} \over X} \eta,\label{a}
\end{equation}
a perfectly legitimate choice, regularizes the integrals in
Eqs.~(\ref{UT}), at least in the $r\to\infty$ region.

The solution (\ref{a}) turns out, as can be explicitely checked using
Eqs.~(\ref{Boxa}) and (\ref{FFa}), to be the only possible non trivial
and asymptotically converging solution. In particular, no dependence
in the string internal coordinates ($z$ and $t$ in our special case)
can be obtained. This means that the simple model used here cannot
lead to current-carrying cosmic
strings~\cite{witten}~\cite{peter}. Moreover, the stationnary solution
(\ref{a}) shows the axion gradient to be orthogonal to $\widetilde
F^{\mu\nu}$, i.e., $\partial_\mu a
\widetilde F^{\mu\nu}=0$. Therefore, Eqs.~(\ref{Boxa}-\ref{BoxA2})
reduce to the usual Nielsen-Olesen set of equations~\cite{NO}, with
the axion coupling using the string solution as a source term. It is
therefore not surprising that the resulting string turns out to be
local.

The total energy per unit length (and tension) is however not finite
in this simple string model for it contains the term
\begin{equation} U = \hbox{f.p.} + 2 \pi \int {dr\over r}
({\delta_{_{GS}} M_{_P} n\over X} - \delta_{_{GS}} M_{_P} A_\theta
)^2,\end{equation} (f.p. denoting the finite part of the integral) so
that, since $A_\theta$ must vanish by symmetry in the string core, one
ends up with
\begin{equation} U = \hbox{f.p.} + 2 \pi  ({\delta_{_{GS}}
M_{_P} n\over X} )^2 \ln ({R_A\over r_a}),\label{div}\end{equation}
where $R_A$ is the radius at which $A_\mu$ reaches its asymptotic
behaviour, i.e., roughly its Compton wavelength $m_{_{A}}$ given in
(\ref{mA}), while $r_a$ is defined as the radius at which the
effective field theory (\ref{lag}) ceases to be valid, presumably of
order $M_{_P}^{-1}$; the correction factor is thence expected to be of
order unity for most theories. Hence, as claimed, the strings in this
model can be made local with a logarithmically divergent energy. The
regularization scale $r_a$ is however a short distance cut-off, solely
dependent on the microscopic structure and does not involve neither
the interstring distance nor its curvature radius. In particular, the
gravitational properties of the corresponding strings are those of a
usual Kibble-Vilenkin string~\cite{vilenkin}, given the equation of
state is that of the Goto-Nambu string $U=T=$const., and the light
deflection is independent of the impact parameter~\cite{defl}.

\section{Local string genesis}

Forming cosmic strings during a phase transition is a very complicated
problem involving thermal and quantum phase
fluctuations~\cite{zurek}. As it is far from being clear how will $a$
and $\eta$ fluctuations be correlated (even though they presumably
will), one can consider to begin with the possibility that a network
of two different kinds of strings will be formed right after the phase
transition, call them $a-$strings and $\eta-$strings, with the meaning
that an $a-$string is generated whenever the axion field winds
(ordinary axion string) while an $\eta-$string appears when the Higgs
field $\Phi$ winds. Both kinds of strings are initially global since
for both of them, only part of the covariant derivatives can be made
to vanish.
We however expect the string network to consist, after some time, in
only these local strings together with the usual global axionic
strings.

Let us consider an axionic string with no Higgs winding: as
$A^\mu\not= 0$, the vacuum solution $\Phi =\rho$ [Eq.~(\ref{12})] is
not a solution, and thus the axionic string field configuration is
unstable. As a result of Eq.~(\ref{Boxphi}), the Higgs field amplitude
tends to vanish in the string core. At this point, it becomes, near
the core, topologically possible for its phase to start winding around
the string, which it will do since this minimizes the total energy
while satisfying the topological requirement that $A_\mu$ flux be
quantized. Such a winding will propagate away from the string.

Conversely, consider the stability of an $\eta-$string with $a=0$. The
conservation of ${\cal J}$ implies, as one can fix $\partial_\mu
A^\mu=0$, that $\Box a =0$, whose general time-dependent solution is
$a = a(|{\bf r}|\pm t)$. Given the cylindrical symmetry, this solution
can be further separated into $a=f(r-t) \theta$. This means that
having a winding of $a$ that sets up propagating away from the string
is among the solutions. As this configuration ultimately would
minimize the total energy, provided $\lim
_{t\to\infty}f=-\delta_{_{GS}}M_{_P}/2X$, this means that the original
string is again unstable and will evolve into the stationary solution
that we derived in the previous section. Note finally that if, instead
of considering the variables $a$ and $\eta$, one had decided to treat
the formation problem by means of the new dynamical variables $\hat a$
and $\ell$, then it would have been clear from the outset that the
resulting string configurations could consist in two different
categories, namely global axionic strings with a winding of $\hat a$,
and local strings with $\ell$ winding.

It should be remarked at this point that these time evolution can in
fact only be accelerated when one takes into account the coupling
between $a$ and $\eta$: if any one of them is winding, then the other
one will exhibit a tendency to also wind, in order to locally minimize
the energy density. Indeed, it is not even really clear whether the
string configurations we started with would even be present at the
string forming phase transition. What is clear, however, is that after
some time, all the string network would consist of local strings
having no long distance interactions. This means in particular that
the relevant scale, if no inflationary period is to occur after the
string formations, should not exceed the GUT scale in order to avoid
cosmological contradictions.

\section*{Conclusions}

Spontaneous breaking of a pseudo-anomalous U(1) gauge symmetry leads
to the formation of cosmic strings whose energy per unit length is
localized around their cores, contrary to what the presence of the
axion field in these theories might have suggested. This happens in
the simple case we've considered here, namely that of a single scalar
field aquiring a VEV at the symmetry breaking. In order to be general,
this result should be generalized to the case where more than one
field gets a VEV; this we now prove.

The potential we consider is that given by Eq.~(\ref{pot}) which, in
full generality, can be rewritten in the form
\begin{equation} V(\Phi_i) = {g^2\over 2} (\Phi ^\dagger {\bf X} \Phi
+ \delta_{_{GS}} M_{_{P}}^2 )^2,\label{pot-gen}\end{equation} where
${\bf X}$ is an $N\times N$ hermitian matrix, and $\Phi$ takes values
in an $N-$dimensional vector space ${\cal V}$. We denote by $p$ the
number of negative eigenvalues of ${\bf X}$ and $\phi$ the restriction
of $\Phi$ to that subspace ${\cal V}_p\in {\cal V}$ spanned by the
eigenvectors of ${\bf X}$ with negative eigenvalues. Once
diagonalized, ${\bf X}$ can be written as
\begin{equation} {\bf X} = \left( \matrix{ {\bf M} & {\bf 0}\cr
{\bf 0} & {\bf P}}\right),\end{equation}
with ${\bf M}$ and ${\bf P}$ containing respectively the negative and
positive eigenvalues.

Eq.~(\ref{pot-gen}) admits an accidental U($N$) symmetry, of which the
anomalous U(1) is part; this is not a simple U($N$)$\times$U(1)
symmetry as each field component transforms differently under U(1) as
indicated on Eq.~(\ref{gauge}). The vacuum configuration is now given
by
\begin{equation} \langle \phi^\dagger {\bf M} \phi \rangle = -
\delta_{_{GS}}
M_{_{P}}^2 ,\end{equation} so it would seem that the remnant symmetry
would be U($N-p$), i.e. a scheme U($N)\to$U($N-p$), and a
topologically trivial vacuum manifold~\cite{kibble}. Hence, one would
naively not expect cosmic string formation in such a model. This
conclusion is in fact not correct, as only part of the original U($N$)
is gauged, namely the anomalous U(1) subgroup, leading to a N\oe{t}her
current
\begin{equation} J^\mu \propto ig [\phi^\dagger {\bf M} \partial^\mu
\phi - (\partial ^\mu\phi^\dagger) {\bf M} \phi ] + 2g^2 A^\mu \phi^\dagger
{\bf M} \phi,\end{equation} which can be made nonzero by imposing a
phase variation for $\phi$ as $\sim \exp (in\theta)$. Once set to a
nonzero value, this current will remain so for topological
reasons~\cite{semi}, being called a semi-local or embedded defect.
All the previous discussions concerning the simple $p=1$ model hold
also for these vortices, including their coupling to the axion field.

The cosmological evolution of the network of strings formed in these
theories also leads to serious constraints on the Green-Schwarz
coefficient provided no domain wall form connecting the strings;
otherwise, the network is known to rapidely (i.e. in less than a
Hubble time) decay into massive radiation and the usual constraint
relative to the axion mass would hold~\cite{kibble,wall-string}. If
however the string network is considered essentially stable, then its
impact on the microwave background limits the symmetry breaking scale
$\delta _{_{GS}} M_{_{P}}$ through the observational requirement that
the temperature fluctuations be not too large~\cite{dT-T}, i.e.  $$GU
\alt 10^{-6},$$ with $G$ the Newton constant $G=
M_{_{P}}^{-2}/(8\pi^2)$. Therefore, the cosmological constraint reads
\begin{equation}
\delta_{_{GS}}  \alt 10^{-2}, \label{delcons}
\end{equation}
a very restrictive constraint indeed. It should be remarked at this
point that the usual domain wall formation leading to a rapid
evaporation of the network does not seem valid for the string solution
we considered. In fact, the axion $\hat a$, defined through
Eq.~(\ref{hata}), as the solution (\ref{a}) set up, vanishes
everywhere. Therefore, when the Peccei-Quinn symmetry is broken, the
axion itself does not have to wind around these strings (it'll do so
however around the ordinary axionic strings present in the
model). Hence, it has no particular reason for taking values in all
its allowed vacuum manifold so that no domain wall will form.

The strings that we have discussed here might appear in connection
with a scenario of inflation. Indeed, the potential (\ref{pot}) is
used for inflation in the scenario known as $D$-term
inflation~\cite{Dterm}: inflation takes place in a direction neutral
under $U(1)$ and the corresponding vacuum energy is simply given by:
\begin{equation}
V_0 = {1 \over 2} g^2 \delta^2_{_{GS}} M_{_P}^4.
\end{equation}
The $U(1)$-breaking minimum is reached after inflation, which leads to
cosmic strings formation. Such an inflation era cannot therefore
dilute the density of cosmic strings and one must study a mixed
scenario~\cite{jeannerot}.  It is interesting to note that, under the
assumption that microwave background anisotropies are predominantly
produced by inflation, the experimental data puts a
constraint~\cite{ERR,LR,KM-R} on the scale $\xi \equiv
\delta_{_{GS}}^{1/2} M_{_P}$ which is stronger than
(\ref{delcons}). Several ways have been proposed~\cite{ERR,KM-R} in
order to lower this scale. They would at the same time ease the
constraint (\ref{delcons}).

A final remark concerning currents is in order at this point. The
strings whose structure we have investigated here are expected to be
coupled to many fields, fermionic in particular. It is well known that
such couplings yield the possibility for the fermions present in the
theory to condense in the string core in the form of so-called zero
modes~\cite{jackiw} which, upon quantization, give rise to
superconducting currents~\cite{witten}. (Note indeed the presence of
anomalies along the worldsheet which was suggested to imply current
inflows~\cite{harvey}.)  Besides, currents tend to raise the
stress-energy tensor degeneracy in such a way that the energy per unit
length and tension become dynamical variables. For loop solutions,
this means a whole new class of equilibrium solutions, named vortons,
whose stability would imply a cosmological
catastrophe~\cite{vortons}. If these objects were to form,
Eq.~(\ref{delcons}) would change into a drastically stronger
constraint. Issues such as whether supersymmetry breaking might
destabilize the currents~\cite{sdad}, thereby effectively curing the
model from the vorton problem, still deserve investigation.

\end{document}